\documentclass[a4paper,10pt,twoside]{cpc-hepnp}
\usepackage{CJK,upgreek,fancyhdr}
\usepackage{multicol}
\usepackage{graphicx}
\usepackage{booktabs}
\usepackage{amssymb,bm,mathrsfs,bbm,amscd}
\usepackage[tbtags]{amsmath}
\usepackage{lastpage}

\usepackage{epstopdf}
\usepackage[colorlinks,linkcolor=blue,anchorcolor=blue,citecolor=blue]{hyperref}
\usepackage{lineno}

\newcommand{\BR}{{\cal B}}
\newcommand{\jpsi}{J/\psi}
\newcommand{\pio}{\pi^{0}}
\newcommand{\pip}{\pi^{+}}
\newcommand{\pim}{\pi^{-}}
\begin{document}
\begin{CJK*}{GB}{gbsn}

\fancyhead[c]{\small Chinese Physics C~~~Vol. 42, No. 1 (2018) 013001}
\fancyfoot[C]{\small 013001-\thepage}

\footnotetext[0]{Received 1 April 2017, Revised 29 July 2017}

\title{Event generators for $\eta/\eta^{\prime}$ decays at BESIII\thanks{Supported by National Natural Science Foundation of China (NSFC) (11205117, 11575133, 11675184), the China Scholarship Council (201506275156), the Wuhan University PhD Short-time Mobility Program 2016, the Joint Funds of the NSFC and Henan Province (U1504112)}}

\author{
 Nian Qin(ñûÄé)$^{1}$%
\quad Zhen-Yu Zhang(ÕÅÕñÓî)$^{1;1)}$\email{zhenyuzhang@whu.edu.cn}%
\quad Shuang-Shi Fang(·¿Ë«ÊÀ)$^{2,3}$\\
\quad Xiang Zhou(ÖÜÏê)$^{1}$
\quad Lin-Lin Du(¶ÅÁØÁØ)$^{4}$
\quad Hao-Xue Qiao(ÇǺÀѧ)$^{1}$
}
\maketitle

\address{
$^1$ Hubei Nuclear Solid Physics Key Laboratory, Key Laboratory of Artificial Micro- and Nano-structures of Ministry of Education, and School of Physics and Technology, Wuhan University, Wuhan 430072, China \\
$^2$ Institute of High Energy Physics, Chinese Academy of Science, Beijing 100049, China \\
$^3$ University of Chinese Academy of Sciences, Beijing 100049, China\\
$^4$ College of Phyics and Materials, Henan Normal University, Xinxiang 453000, China \\
}
%\linenumbers
\begin{abstract}
The light unflavoured meson $\eta/\eta^{\prime}$ decays are valuable for testing non-perturbative quantum chromodynamics and exploring new physics beyond the Standard Model. This paper describes a series of event generators, including
$\eta/\eta^{\prime}\to\gamma l^{+}l^{-}$, $\eta/\eta^{\prime}\to\gamma \pi^{+}\pi^{-}$, $\eta^{\prime}\to\omega e^{+}e^{-}$, $\eta\to\pi^{+}\pi^{-}\pi^{0}$, $\eta/\eta'\to\pio\pio\pio$, $\eta^{\prime}\to\eta\pi\pi$ and $\eta'\to\pip\pim\pip\pim/\pip\pim\pio\pio$, which have been developed for investigating $\eta/\eta^\prime$ decay dynamics. For most of these generators, their usability has been validated in BESIII analyses for determining the detection efficiency, and background studies. The consistency between data and Monte Carlo shows that these generators work well in the BESIII simulation, and will also be useful for ongoing BESIII  analyses and other experiments for studying $\eta/\eta^\prime$ physics.
\end{abstract}



\footnotetext[0]{\hspace*{-3mm}\raisebox{0.3ex}{$\scriptstyle\copyright$}Content from this work may be used under the terms of the Creative Commons Attribution 3.0 licence. Any further distribution of this work must maintain attribution to the author(s) and the title of the work, journal citation and DOI. Article funded by SCOAP3 and published under licence by Chinese Physical Society and the Institute of High Energy Physics of the Chinese Academy of Sciences and the Institute of Modern Physics of the Chinese Academy of Sciences and IOP Publishing Ltd}%

\begin{multicols}{2}

\section{Introduction}
As the ground states of pseudoscalar nonets, the $\eta$ and $\eta^\prime$ mesons have been firmly established and their main decay modes are fairly well known~\cite{PDG2014}. How\-ever, they still attract theoretical and experimental attention due to their special role in understanding low energy quantum chromodynamics (QCD), even though they were discovered about half a century ago~\cite{eta_discovery}.
The decays of $\eta/\eta^\prime$ are of interest as probes of some aspects of strong interactions, and also as sources of information on physics beyond the Standard Model (SM).

Due to flavor symmetry breaking, $\eta/\eta'$ mesons involve the mixing of an octet state and a singlet state, with a mixing angle of about $-20^{\circ}$~\cite{mixing_eta}.  A larger probability to be a singlet state gives a larger mass to the $\eta'$, because the axial anomaly only contributes to the singlet mass.
The gluonic content in the $\eta'$, which is related to vacuum topology and the U(1) anomaly~\cite{QCD_anomaly}, may also contribute to its large mass.

  The unflavored  $\eta/\eta'$ mesons not only play an important role in studying the interactions between light quarks and the interactions between quarks and glu\-ons~\cite{Gluonic_meson}, but also offer  a unique place to test fundamental symmetries in QCD in the low energy region.  In addition, their hadronic decays could be used to determine the difference of light quark masses~\cite{mass_diff}. Therefore, $\eta/\eta^\prime$ physics is listed in the programs of many experiments, such as BESIII, KLOE-2, MAMI, GlueX, and CLAS12.  Most recently, a new facility, REDTOP~\cite{REDTOP_ex}, is also proposed to study $\eta/\eta^\prime$ decays.

Due to the large production rate of $\eta/\eta^\prime$ mesons in $J/\psi$ hadronic and radiative decays, the world's largest sample of $1.31\times 10^9$ $\jpsi$ events~\cite{JpsiNumber}, collected with the BESIII detector, offers a good opportunity to study the decays of $\eta/\eta^{\prime}$. In recent years BESIII has achieved much progress on $\eta/\eta^{\prime}$ decays~\cite{etap_pipiee_bes, etap_gammaee_bes, eta_3pi_bes, etap_4pi_bes, etap_omegaee_bes, etap_Kpi_bes}. Experimentally, a good description of the amplitude for each decay mode in the Monte Carlo (MC) simulation plays an important role in the optimization of the event selection criteria, determination of the detection efficiency, and background suppression. In the near future,
about 10 billion $J/\psi$ events will have been accumulated at the BESIII detector, which will allow us to search for the rare decays of $\eta/\eta^\prime$ at an unprecedented level. In that case, the established $\eta/\eta^\prime$ decays become the dominant background sources and proper event generators are hence essential to the background estimation.

To meet the challenge of precision measurement of $\eta/\eta^\prime$ physics, in this paper we present a series of event\- generators for  their decays, including  $\eta/\eta^{\prime}\to\gamma l^{+}l^{-}$, $\eta/\eta^{\prime}\to\gamma \pi^{+}\pi^{-}$, $\eta^{\prime}\to\omega e^{+}e^{-}$, $\eta\to\pi^{+}\pi^{-}\pi^{0}$, $\eta/\eta'\to\pio\pio\pio$, $\eta^{\prime}\to\eta\pi\pi$ and $\eta'\to\pip\pim\pip\pim/\pip\pim\pio\pio$, within the framework of chiral perturbation theory (ChPT) and the vector meson dominance (VMD) model. In Section 2, a brief introduction is given to describe the software framework for event generators. In Section 3, the decay amplitudes for these decays and the corresponding parameters used for generating events are provided.  Meanwhile, validations for some of these generators are also performed to ensure that they work well in
the BESIII MC simulation package.  A short summary is presented in the last section.

%Then several distributions from the simulations are inspected, which are in good agreement with the theoretical predictions, indicating that the event generators work very well with the BESIII Monte Carlo simulation %packages. A condensed summary is presented in the last section.
%to measure a decay or a scattering process, Monte Carlo simulation should be run firstly, of which generator is one of the essential components. It is essential to build dynamic generator for the rare decays %of $\eta/\eta'$ in order to simulate the process as close as possible to the physics, improve the detection efficiencies and suppress backgrounds.
%Reference~\cite{zhangzygenerators} has offered reliable event generators for $\eta/\eta^{\prime}\to\pi^{+}\pi^{-}e^{+}e^{-}/\pi^{+}\pi^{-}\mu^{+}\mu^{-}/e^{+}e^{-}\mu^{+}\mu^{-}$, however, more event %generators are required.

%The chiral perturbation theory (ChPT) and vector meson dominance (VMD) model are used to describe the different decay modes of $\eta/\eta'$.

\section{Software framework for event generators}
%Usually the MC simulation is used to determine the mass resolution, detection efficiency and the background contribution in the experimental data analysis.
At the BESIII experiment, the \textsc{geant4}-based simulation
software \textsc{boost} includes the geometric and material
description of the BESIII detector, the detector response, and the
digitization models, as well as the detector running conditions and
performance~\cite{ref:boost}.  The charmonium state, $e.g.$, $\jpsi$, is simulated with
the MC event generator \textsc{kkmc}~\cite{ref:kkmc,ref:kkmc2}, while the
decays are generated by BesEvtGen~\cite{ref:evtgen} for known decay modes,
with branching fractions set to the world average values in the Particle Data Group
(PDG)~\cite{PDG2014}, and by
\textsc{lundcharm}~\cite{ref:lundcharm} for the remaining unknown decays.  The
event generators are based on the framework of the BESIII offline software
system (BOSS)~\cite{ref:boss}.

In general, the MC simulation in BOSS is performed by passing the Lorentz-vector
of all the particles produced by the event generator into a simulation package of the
BESIII detector after taking into account the detector construction, the detector
response, the interaction between the particles, and the material~\cite{zhangzygenerators}.
In this paper, the event generators  were developed in accordance with the corresponding $\eta/\eta^\prime$ decay amplitudes, which are described in detail in Section ~\ref{amplitude}, and
then were implemented in BesEvtGen.

\section{Theoretical formulas and simulations}
\label{amplitude}
\subsection{$\eta/\eta^{\prime}\to\gamma l^+l^-$}
Electromagnetic Dalitz decays of  $\eta/\eta^\prime\to\gamma l^+ l^-$ ($l^\pm$ stands for $\mu^\pm$ or $e^\pm$) play an important role in revealing the structure of hadrons and the interaction mechanism between photons and hadrons. This  is the so-called single off-shell decay in which  the $l^+l^-$ pair originates from the off-shell photon ($\gamma^*$).
The four-momenta for the process $\eta/\eta^\prime \to \gamma(k)\gamma^*(p)\to \gamma(k) l^+(p_1)l^-(p_2)$ are defined as $P=k+p=k+p_1+p_2$. The square of the amplitude can be written as~\cite{anomalousdecayofP}
\begin{equation}
  \begin{aligned}
        &|\mathcal{A}(P\to l^+l^-\gamma)|^2\\
        &=e^2|\mathcal{M}_{P}(p^2,k^2=0)|^2 \dfrac{(m_{P}^2-p^2)^2}{2p^2}(2-\beta_p^2\sin^2\theta_p),
  \end{aligned}
\end{equation}
where $p=p_1+p_2$,
$\beta_{p}=\sqrt{1-\dfrac{4m_{l^\pm}^{2}}{p^{2}}}$
and $\theta_p$ is the helicity angle. $\mathcal{M}_{P}(p^2,k^2=0)$ is the form factor, which is described as
\begin{equation}
\mathcal{M}_{P}(p^2,k^2=0)=\mathcal{M}_{P}\times VMD(p^2),
\end{equation}
where
\begin{equation}
\mathcal{M}_{P}=
\left\{
\begin{aligned}
\dfrac{\alpha}{\pi f_{\pi}}\dfrac{1}{\sqrt{3}}\left(\dfrac{f_{\pi}}{f_8}\cos\theta_{mix}-2\sqrt{2}\dfrac{f_{\pi}}{f_0}\sin\theta_{mix}\right) \text{if} ~P=\eta; \\
\dfrac{\alpha}{\pi f_{\pi}}\dfrac{1}{\sqrt{3}}\left(\dfrac{f_{\pi}}{f_8}\sin\theta_{mix}+2\sqrt{2}\dfrac{f_{\pi}}{f_0}\cos\theta_{mix}\right) \text{if} ~P=\eta', \\
\end{aligned}
\right.
\end{equation}
where $\alpha=1/137$, $f_{\pi}=92.4$ MeV, $f_0=1.04f_{\pi}$, $f_8=1.3f_{\pi}$ and $\theta_{mix}=-20^{\circ}$~\cite{parameters_s1}.
The $\eta/\eta'\to\gamma l^+ l^-$ form factor measurements support the theoretical prediction from the VMD model that the dominant contribution is from the $\rho$~\cite{eta2gamLL_ex,etap_gammaee_bes}.
The VMD form factor for $\eta\to\gamma l^+ l^-$ in the generator is written as
\begin{equation}
\label{vmd_eta}
VMD(s)=1-c_3+c_3\dfrac{1}{1-\dfrac{s}{m_\rho^2}-i\dfrac{\Gamma(s)}{m_\rho}},
\end{equation}
where $s$ is defined as $s=p^2=(p_1+p_2)^2$.
$\Gamma(s)$ is the width of the vector meson~\cite{width_rho}
\begin{equation}
\label{eq5_gamLL}
\Gamma(s)=\Gamma_\rho\dfrac{\sqrt{s}}{m_\rho}\left(\dfrac{1-\dfrac{4m_{\pi}^{2}}{s}}{1-\dfrac{4m_{\pi}^{2}}{m^{2}_\rho}}\right)^{3/2}
\Theta(s-4m_{\pi}^2),
\end{equation}
and the $\Theta$ function is
\begin{equation}
\Theta(s-4m_{\pi}^2)
=
\left\{
\begin{aligned}
1,~~\text{if}  ~s\geq4m_{\pi}^2 ; \\
0,~~\text{if}  ~s<4m_{\pi}^2 , \\
\end{aligned}
\right.
\end{equation}

For $\eta^\prime \rightarrow\gamma l^+l^-$, the phase space allows production of $\rho$ and $\omega$. Even although the contribution from $\omega$ is small, it cannot be directly ignored,
since its interference with $\rho$ may lead to a sizable contribution.
%For the generators of $\eta'\to\gamma l^+ l^-$, there is not only a contribution from the $\rho$, but also from the $\omega$.
By combining with $\rho$ and $\omega$, the VMD form factor for $\eta'\to\gamma l^+ l^-$ can be described by
\begin{equation}
\label{vmd_etap}
VMD(s)=\dfrac{w_{\rho}\cdot BW_{\rho} + w_{\omega}\cdot BW_{\omega}}{(w_{\rho}+w_{\omega})},
\end{equation}
where $BW_{\rho}(BW_{\omega})$ represents a simple Breit-Wigner function for $\rho(\omega)$.

In the generator, the parameters in the above formula are set to be $m_\rho=775.49$ MeV, $m_{\pi}=139.57$ MeV,
and $\Gamma_\rho=149.1$ MeV~\cite{PDG2014}. The weight factors $w_{\rho}$ and $w_{\omega}$ are subjected to $w_{\rho}:w_{\omega}=3:1$~\cite{weight_rho_omega}.
The different vector meson dominance models can be switched by inserting different values of $c_3$~\cite{CouplingConstant1, CouplingConstant2}.

By implementing the above amplitudes in the BESIII simulation package, the $\eta/\eta^{\prime}\to\gamma l^+l^-$ events are generated. The mass spectra of the leptonic pairs (the solid histograms) at the truth level are shown in Fig.~\ref{GamEE_draw}, where the VMD form factor used for generating $\eta\rightarrow\gamma l^+l^-$  events is that for the case of the hidden gauge ($c_3=1$). For a comparison, we also generate the events
with $VMD(s)=1$, as indicated by the hatched histograms in Fig.~\ref{GamEE_draw}.  The discrepancies between the events generated with and without VDM form factors are quite obvious,  in particular for
$\eta/\eta^\prime\to\gamma\mu^+\mu^-$, which also shows that reliable dynamic generators are very important for the study of $\eta/\eta'$ physics.

%  the mass spectra, $M(l^+l^-)$, for  $\eta\to\gamma e^+e^-$ and $\eta\to\gamma \mu^+\mu^-$ were displayed  as the solid histograms in Fig.~\ref{GamEE_draw} (a) and Fig.~\ref{GamEE_draw} (c).

 %simulated at the truth level, which the detector response, the detector construction, and the interaction between the particles and the material are ignored.
%The DIY event generators are performed in the hidden gauge case ($c_3=1$) with the VMD factor in Eq.(\ref{vmd_eta}) for $\eta\to\gamma l^+l^-$.
%The VMD factor in Eq.(\ref{vmd_etap}) is used in the generators for $\eta'\to\gamma l^+l^-$.
%The invariant masses of the dileptons obtained from the MC simulations are shown in Fig.~\ref{GamEE_draw}.
%{\color{blue}
%The solid histograms are simulated with the VMD form factors described in the text. The shaded histograms are from the simulations with the VMD form factors set to be 1.0. The discrepancies between the solid histograms and the shaded histograms are obvious in the decays of $\eta/\eta'\to\gamma\mu^+\mu^-$, which indicate that the reliable dynamic generators are very important for the study of $\eta/\eta'$ physics.
%The typical strong peaks around the small asses emerge cl in Figs.~\ref{GamEE_draw} (a) and (b.1), while there are also peaks around 770 MeV shown as solid histograms in Figs.~\ref{GamEE_draw} (b.2) and (d), %indicating the contributions from the vector mesons.
\end{multicols}
%\ruleup
\begin{center}
\includegraphics[width=14cm]{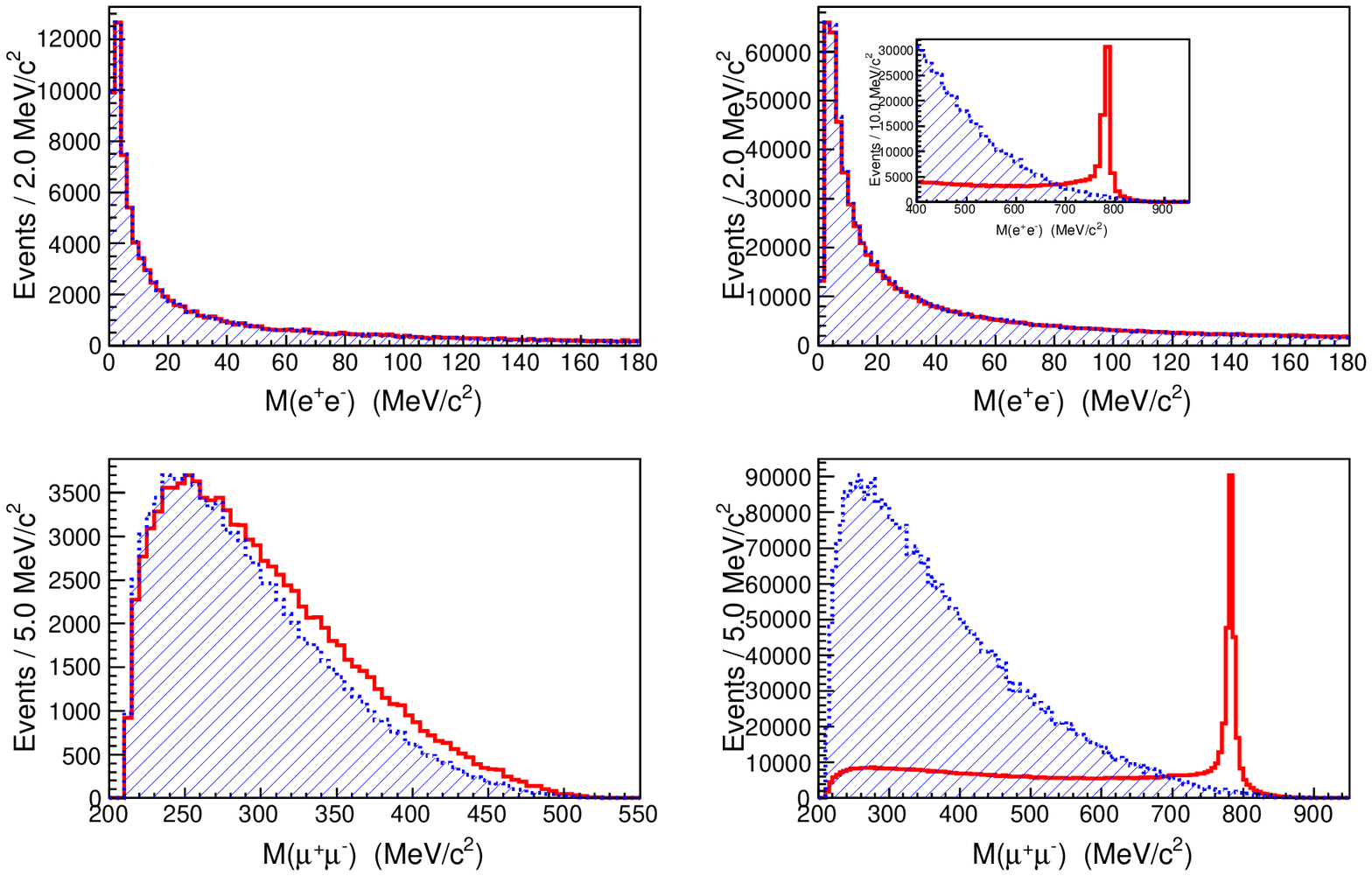}\put(-230,230){\bf (a)}\put(-40,230){\bf (b.1)}\put(-120,220){\bf (b.2)}\put(-230,100){\bf (c)}\put(-30,100){\bf (d)}
\figcaption{\label{GamEE_draw} The invariant mass distributions obtained from simulations of the DIY generators: (a) $M(e^+e^-)$ from $\eta\to\gamma e^+ e^-$; (b.1) and (b.2) M$(e^+e^-)$  from $\eta'\to\gamma e^+ e^-$; (c) $M(\mu^+\mu^-)$ from $\eta\to\gamma \mu^+ \mu^-$; (d) $M(\mu^+\mu^-)$ from $\eta^\prime\to\gamma \mu^+ \mu^-$. The solid histograms are from simulations with the VMD form factors described in the text, and the shaded histograms are from simulations with $VMD(s)=1$.}

%{\color{blue}
%The solid histograms are from the simulations with the VMD form factors described in the text. The shaded histograms are from the simulations with the VMD form factors set to be 1.0.
%}
%(a) is from $\eta\to\gamma e^+ e^-$; (b.1) and (b.2) are from $\eta'\to\gamma e^+ e^-$; (c) is from $\eta\to\gamma \mu^+ \mu^-$ and (d) is from $\eta'\to\gamma \mu^+ \mu^-$.}
\end{center}
%\ruledown

\begin{multicols}{2}

\subsection{$\eta/\eta^{\prime}\to\gamma\pip\pim$}
The decay $\eta/\eta^{\prime}\to\gamma\pip\pim$ receives a contribution from the box anomaly~\cite{anomalousdecayofP}. For the $\eta\to\gamma(k)\pip(p_1)\pim(p_2)$ decay, the unpolarized squared decay amplitude is ~\cite{anomalousdecayofP}
\begin{equation}
\label{eq7_gampipi}
  \begin{aligned}
\sum^2_{pol=1}|\mathcal{A}(\eta\to\gamma\pip\pim)|^2(s, \theta_p)=
&\dfrac{\lambda(m^2_\eta, s, 0)s\beta^2_p\sin^2\theta_p}{16m^6_\eta}\\
&\cdot\left(|M_G|^2+|E_G|^2\right),
  \end{aligned}
\end{equation}
where  $s=p^2=(p_1+p_2)^2$, $\beta_p=\sqrt{1-\dfrac{4m_{\pi^\pm}^{2}}{p^2}}$, and $\theta_p$ is the polar angle of $p_{\pi^\pm}$ in the $p_{\pi^+}~p_{\pi^-}$ rest frame with respect to the direction of the flight of the $p_{\pi^+}~p_{\pi^-}$ in the rest frame of the pseudoscalar meson.
The electric form factor $E_G$ is used to describe the CP violation in the decay, which is set to be zero in this paper.
The K{\"a}ll{\'e}n function is
\begin{equation}
\lambda(x, y, z)=x^2+y^2+z^2-2xy-2yz-2xz,
\end{equation}
where it is $\lambda(m^2_\eta, s, 0)=(m^2_\eta-s)^2$ in our DIY generator.
The magnetic form factor is $M_G(s)=m_\eta^3\mathcal{M}_\eta(s)$, when
\begin{equation}
\mathcal{M}_\eta(s)=\mathcal{M}_\eta\times VMD(s),
\end{equation}
with
\begin{equation}
\mathcal{M}_\eta=\dfrac{e}{8\pi^2f^3_\pi}\dfrac{1}{\sqrt3}\left(\dfrac{f_\pi}{f_8}\cos\theta_{mix}-2\sqrt2\dfrac{f_\pi}{f_0}\sin\theta_{mix}\right),
\end{equation}
and
\begin{equation}
\label{eq11_gampipi}
VMD(s)=1-\dfrac{3}{2}c_3+\dfrac{3}{2}c_3\dfrac{m^2_\rho}{m^2_\rho-s-im_\rho\Gamma(s)},
\end{equation}
where $\Gamma(s)$ is the same as in Eq.~(\ref{eq5_gamLL}).

At first, the decay $\eta'\to\gamma\pip\pim$ was believed to be dominated  by $\eta'\to\gamma\rho$ with the subsequent decay $\rho\to\pip\pim$. In this case, the squared amplitude of $\eta'\to\gamma\pip\pim$ would be similar to Eq.~(\ref{eq7_gampipi}). However, the $\rho$ mass extracted from the dipion mass spectrum by different experiments is about 20 MeV/$c^2$ larger than that from $e^+e^-$ annihilation~\cite{etap2gampipi_mesure}. This effect is accounted for by the higher term of the Wess-Zumino-Witten (WZW) ChPT Lagrangian describing the non-resonant part of the coupling~\cite{etap2gampipi_WZW}. A simple Breit-Wigner function for $\rho$ in the form factor in Eq.~(\ref{eq11_gampipi})  is not enough to describe the data.
The $\rho-\omega$ interference and the box anomaly should be taken into account for $\eta'\to\gamma\pip\pim$.
Deduced from the ones used in Ref.~\cite{etap2gampipi_th}, the decay rate~\cite{etap_pipiee_bes} can be expressed by
\begin{equation}
\label{eq12_gampipi}
\dfrac{d\Gamma}{dm}\propto k^3_\gamma q^3_\pi(m)|\text{BW}^{\text{GS}}_\rho(1+\delta\dfrac{m^2}{m^2_\rho}\text{BW}_\omega)+\beta|^2,
\end{equation}
where
$m^2=(p_{\pi^+}+p_{\pi^-})^2$, and $p_{\pi^+}$ and $p_{\pi^-}$ are the four-momenta in the laboratory frame.
$k_\gamma$ is the photon energy and $q_\pi(m)$ is the momentum of the pion in the $\pip\pim$ rest frame.
The parameter $\delta$ is a complex number, for which $|\delta|=5.59\times10^{-4}$ represents the contribution from the $\omega$ resonance and the complex phase of $\delta$ (arg$\delta=-3.78$ rad) represents the interference between the $\omega$ and the $\rho(770)$ resonance~\cite{etap_pipiee_bes}.
$m_\rho$ is the mass of the $\rho(770)$ resonance.
$\beta=-19.33$ is the box anomaly constant ratio, which represents the non-resonant contribution~\cite{etap_pipiee_bes}.
$\text{BW}_\omega$ represents a simple Breit-Wigner function for $\omega$. $\text{BW}^{\text{GS}}_\rho$ is the Breit-Wigner distribution in the GS parametrization~\cite{etap2gampipi_GS},
\begin{equation}
\label{eq13_BWgs}
\text{BW}^{\text{GS}}_\rho=\dfrac{m^2_\rho(1+d\cdot\Gamma_\rho/m_\rho)}{m^2_\rho-m^2+f(m^2)-im_\rho\Gamma_\rho(m)},
\end{equation}
where
\begin{equation}
  \begin{aligned}
f(m^2)=&\Gamma_\rho\dfrac{m^2_\rho}{q^3_\pi(m^2_\rho)}[q^2_\pi(m^2)\cdot(h(m^2)-h(m^2_\rho))+(m^2_\rho-m^2)\\
       &\cdot q^2_\pi(m^2_\rho)\cdot\dfrac{dh}{dm^2}|_{m^2=m^2_\rho},
  \end{aligned}
\end{equation}
and $\dfrac{dh}{dm^2}|_{m^2=m^2_\rho}$ is
\begin{equation}
\dfrac{dh}{dm^2}|_{m^2=m^2_\rho}=h(m^2_\rho)[(8q^2(m^2_\rho))^{-1}-(2m^2_\rho)^{-1}]+(2\pi m^2_\rho)^{-1}.
\end{equation}
The function $h(m^2)$ is defined as
\begin{equation}
h(m^2)=\dfrac{2}{\pi}\dfrac{q_\pi(m^2)}{m}ln\dfrac{m+2q_\pi(m^2)}{2m_\pi},
\end{equation}
when
\begin{equation}
d=\dfrac{3}{\pi}\dfrac{m^2_\pi}{q^2_\pi(m^2_\rho)}ln\dfrac{m_\rho+2q_\pi(m^2_\rho)}{2m_\pi}+\dfrac{m_\rho}{2\pi q_\pi(m^2_\rho)-\dfrac{m^2_\pi m_\rho}{\pi q^3_\pi(m^2_\rho)}},
\end{equation}
where $q_\pi(m^2)=\sqrt{m^2/4-m^2_\pi}$ is the momentum of the pion in the $\pip\pim$ rest frame with $m_\pi=139.57$~MeV~\cite{PDG2014}, $q_\pi(m^2_\rho)=\sqrt{m^2_\rho/4-m^2_\pi}$ is the momentum of the pion in the $\pip\pim$ rest frame with $m=m_\rho$, and $\Gamma_\rho(m)=\Gamma_\rho(\dfrac{q_\pi(m^2)}{q_\pi(m^2_\rho)})^3(\dfrac{m_\rho}{m})$.

At the truth level, Figs.~\ref{gampipi_draw}(a) and (b) show the $\pi^+\pi^-$ mass spectra for $\eta\rightarrow\gamma\pi^+\pi^-$ and $\eta^\prime\rightarrow\gamma\pi^+\pi^-$, respectively.  At  the detector level, the generator
for $\eta\to\gamma\pi^+\pi^-$ has been validated in the BESIII measurement of $\eta'\to4\pi$~\cite{etap_4pi_bes} by providing a good description of the background events from $\eta^\prime\rightarrow\pi^+\pi^-\eta (\eta\rightarrow\gamma\pi^+\pi^-)$.  The generator for $\eta^\prime\rightarrow\gamma\pi^+\pi^-$ has also been used in the determination of the detection efficiency of $J/\psi\rightarrow\gamma\eta^\prime (\eta^\prime\rightarrow\gamma\pi^+\pi^-)$ at the BESIII experiment. It was found that the $\pi^+\pi^-$ mass spectrum from the MC simulation is consistent with that of data.
More model-independent approaches can be found in Refs.~\cite{etap2gampipi_more1, etap2gampipi_more2}.

%The MC samples are generated to check the DIY event generators at the truth level. The distributions from the MC samples are shown in Fig.~\ref{gampipi_draw}.
%{\color{blue}

%The generator of $\eta\to\gamma \pi^+ \pi^-$ is used to describe the backgrounds with the full simulation considering the detector performance in the BESIII analysis of $\eta'\to4\pi$~\cite{etap_4pi_bes}.
%The decay $\eta'\to\gamma \pi^+ \pi^-$ is simulated with the decay rate formula in Eq.~(\ref{eq12_gampipi}) and the GS parameterized Breit-Wigner function in Eq.~(\ref{eq13_BWgs}), taking into account the $\rho-\omega$ %interference and the box anomaly.
%The full simulation of $\eta'\to\gamma \pi^+ \pi^-$ generator is applied in the BESIII analysis of $J/\psi\to\gamma\eta'$ to study the backgrounds and obtain the detection efficiency~\cite{etap_pipiee_bes}.%
%More model-independent approaches can be found in references~\cite{etap2gampipi_more1, etap2gampipi_more2}.
%}

\end{multicols}
%\ruleup
\begin{center}
\includegraphics[width=7cm]{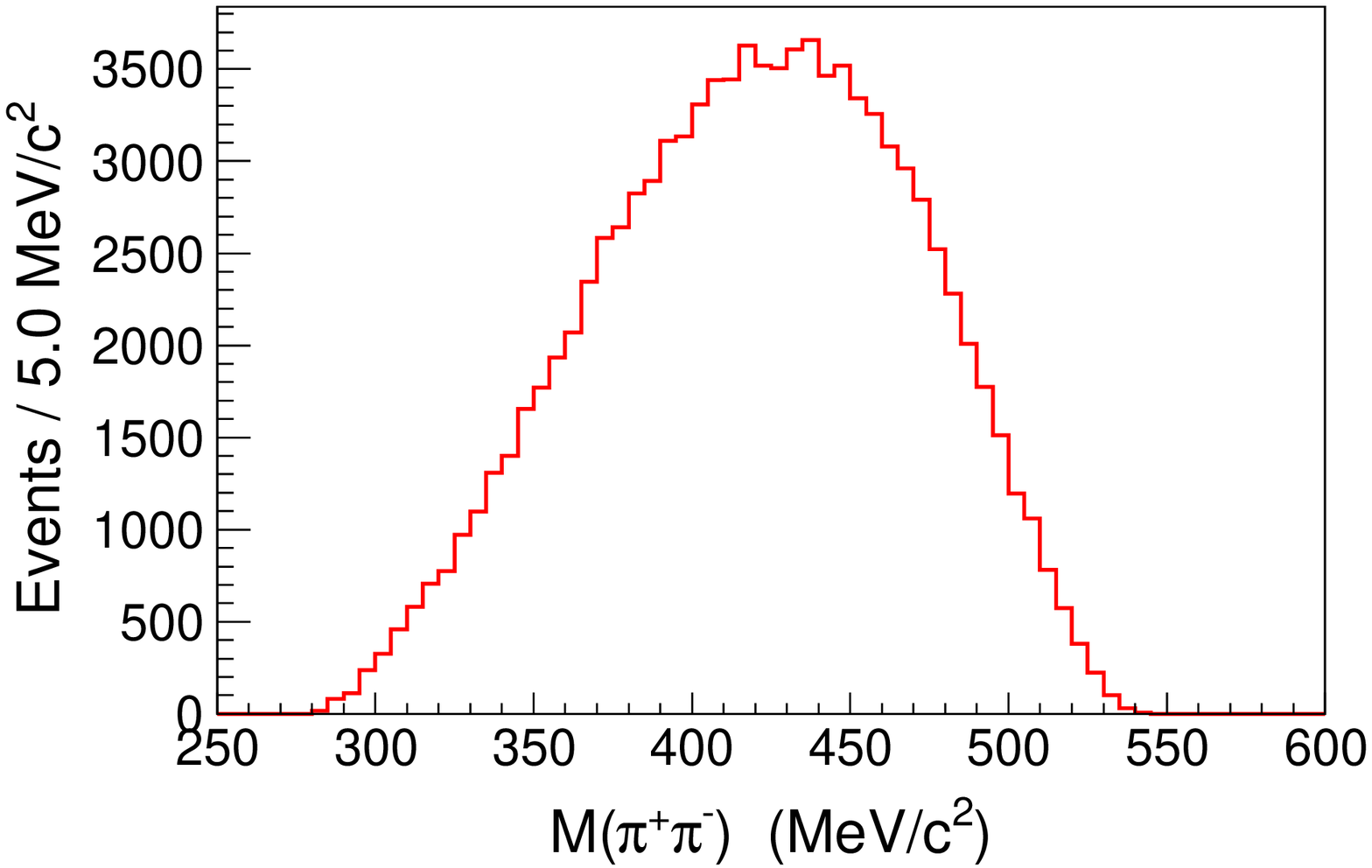}\put(-30,100){\bf (a)}
\includegraphics[width=7cm]{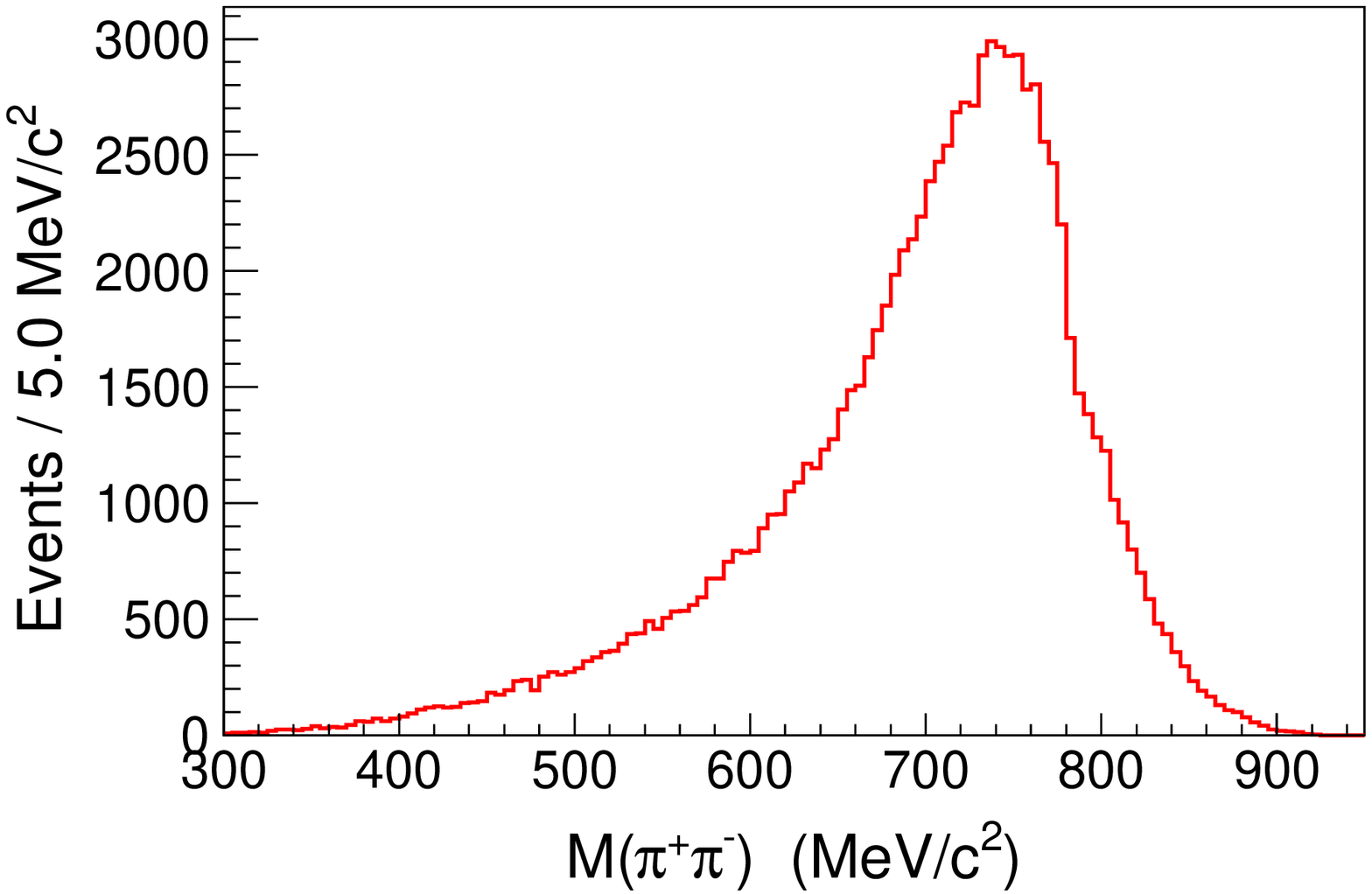}\put(-30,100){\bf (b)}
\figcaption{\label{gampipi_draw} The $\pi^+\pi^-$ invariant mass distributions  for (a) $\eta\to\gamma \pi^+ \pi^-$ and  (b)  $\eta'\to\gamma \pi^+ \pi^-$ at the truth level.}
\end{center}
%\ruledown

\begin{multicols}{2}

\subsection{$\eta'\to\omega e^+e^-$}
It is interesting to study the decay $\eta'\to V e^+e^-$ ($V$ represents a vector meson), which is related to the two-body radiative decay into a vector meson and an off-shell photon. The $e^+e^-$ invariant mass distribution will provide us useful information on the internal structure of the $\eta'$ meson and the momentum dependence of the transition form factor. In 2013, BESIII reported the measurement of $\eta'\to\pip\pim e^+e^-$~\cite{etap_pipiee_bes}, which is found to be dominated by $\eta'\to\rho e^+e^-$,  and the results are in agreement with the theoretical predictions~\cite{etap2pipiee_th1, etap2pipiee_th2}. More recently, the branching fraction  $\BR(\eta'\to\omega e^+e^-)=[1.97\pm0.34\pm0.17]\times10^{-4}$ was reported for the first time via $\jpsi$ radiative decays~\cite{etap_omegaee_bes}, and is also consistent with the theoretical predictions~\cite{etap2pipiee_th1, etap2omegaee_th}.

Within the framework of  effective meson theory~\cite{etap2pipiee_th1}, the square of the amplitude of $\eta'\to\omega e^+e^-$ can be written as
\begin{equation}
  \begin{aligned}
        &|\mathcal{A}(P\to Ve^+e^-)|^2\\
        =&2\pi\alpha\Gamma_{P\to\gamma V}\dfrac{32\pi m^3_P}{(m^2_P-m^2_V)^3}|VMD(p^2)|^2\\
         &\cdot\dfrac{(m^2_P-p^2-m^2_V)^2-4m^2_Vp^2}{p^2}(2-\beta^2_p\sin^2\theta_p)\\
        =&\dfrac{2^6\pi^2m^3_P\alpha\Gamma_{P\to\gamma V}}{(m^2_P-m^2_V)^3}|VMD(p^2)|^2\dfrac{(m^2_P-p^2-m^2_V)^2-4m^2_Vp^2}{p^2}\\
        &\cdot(2-\beta^2_p\sin^2\theta_p),
  \end{aligned}
\end{equation}
where $p^2=(p_{l^+}+p_{l^-})^2$, $p_{l^\pm}$ is the four-momentum of the dilepton;
$\Gamma_{P\to\gamma V}$ is the decay width of $\eta'\to\gamma\omega$;
$m_P$ and $m_V$ are the mass of the pseudoscalar and vector meson respectively in the process  $P\to\gamma V$; $\beta_{p}=\sqrt{1-\dfrac{4m_{l^\pm}^{2}}{p^{2}}}$; and $\theta_p$ is the polar angle of $p_{l^\pm}$ in the $p_{l^+}~p_{l^-}$ rest frame with respect to the direction of flight of the $p_{l^+}~p_{l^-}$ in the pseudoscalar rest frame. The vector meson dominance form factor can be written as
\begin{equation}
%{\color{red} VMD(p^2)=\dfrac{1}{1-\dfrac{p^2}{m_\omega^2}-i\dfrac{\Gamma_\omega}{m_\omega}} },
VMD(p^2)=\dfrac{w_{\omega}\cdot BW_{\omega} + w_{\phi}\cdot BW_{\phi}}{(w_{\omega}+w_{\phi})},
\end{equation}
where $m_\omega(m_\phi)$ and $\Gamma_\omega(\Gamma_\phi)$ are the mass and width, respectively, of $\omega(\phi)$ in the PDG~\cite{PDG2014}.
$BW_{\rho}(BW_{\omega})$ represents a simple Breit-Wigner function for $\rho(\omega)$.
The weight factors $w_\omega$ and $w_\phi$ are subjected to $w_\omega:w_\phi=1:4$~\cite{etap2omegaee_weight}.

Figure~\ref{etap2omegaee_draw} shows the $e^+e^-$ invariant mass distribution at the truth level. In the observation of $\eta'\to e^+e^-\omega$~\cite{etap_omegaee_bes}, the MC events generated with this generator provided a good description of the data.

%\citeThe MC sample is generated to check the DIY event generator at the truth level, which the distribution is shown in Fig.~\ref{etap2omegaee_draw}. The shape of the invariant mass spectrum is consistent with theoretical %result.
%{\color{red}
%The generated $e^+ e^-$ mass spectrum with the full simulation is in good agreement with the BESIII result~\cite{etap_omegaee_bes}.
%}

\begin{center}
  \includegraphics[width=7cm]{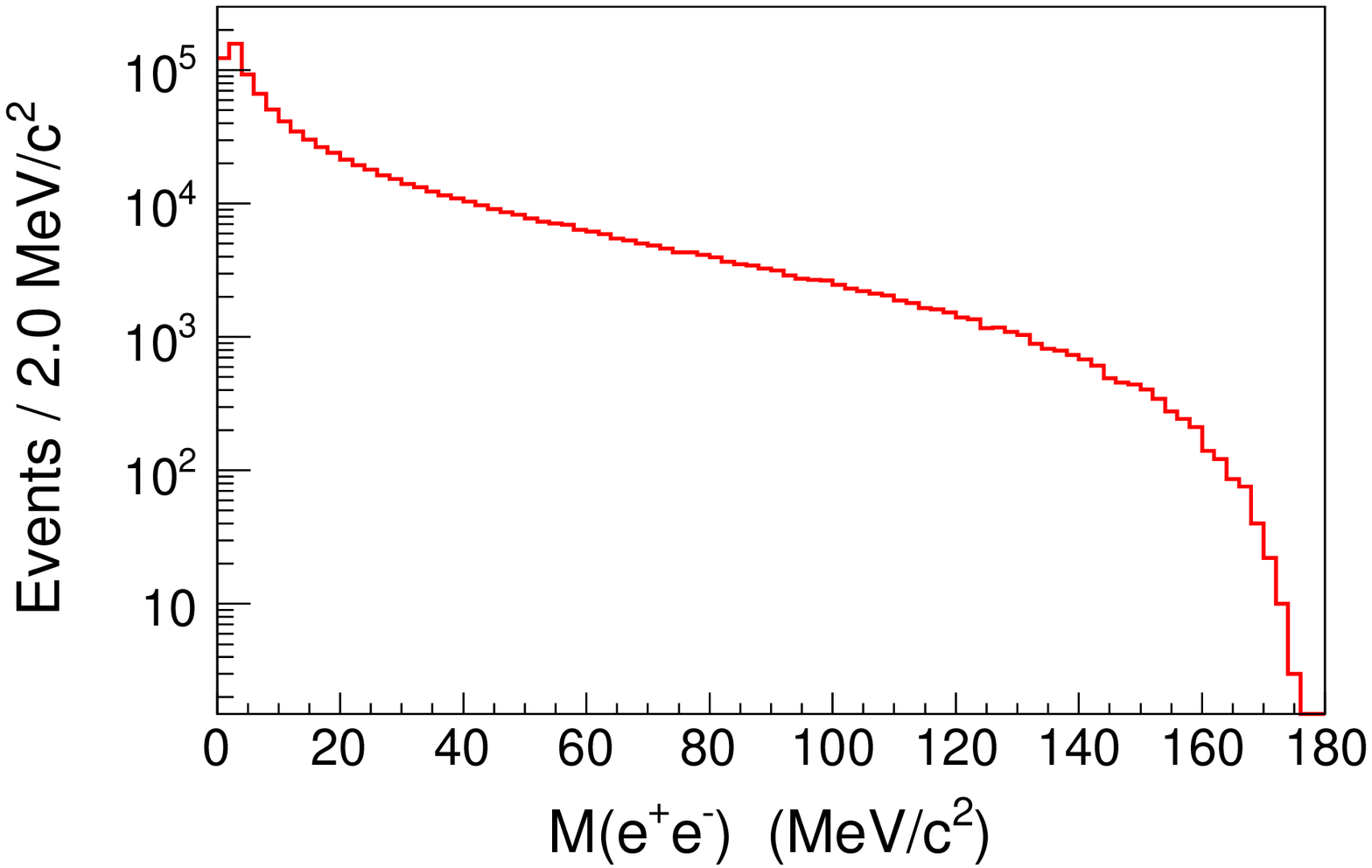}
  \figcaption{\label{etap2omegaee_draw}The  mass spectrum of $e^+e^-$ obtained from the MC simulation for the decay $\eta'\to\omega e^+e^-$.}
\end{center}

\subsection{$\eta\to\pip\pim\pio, \eta/\eta'\to\pio\pio\pio$}
The decays $\eta/\eta'\to 3\pi$ violate G parity and are induced dominantly by the strong interaction via the $u-d$ quark mass difference.
It offers an ideal laboratory for testing ChPT and provides validation of models for the $\pi$-$\pi$ final-state interaction~\cite{eta23pi_th}.
BESIII has measured the branching fractions of $\eta/\eta'\to 3\pi$ for both charged and neutral channels~\cite{eta23pi_ex1}, and a Dalitz plot analysis is also reported~\cite{eta_3pi_bes}.
The internal dynamics of charged decay channel ($\eta\to\pip\pim\pio$) can be described by two independent Dalitz plot variables~\cite{eta23pi_amp}
\begin{equation}
  \begin{aligned}
        &X = \dfrac{\sqrt{3}}{Q}(T_{\pip}-T_{\pim}),\\
        &Y = \dfrac{3T_{\pio}}{Q}-1,
  \end{aligned}
\end{equation}
where $T_\pi$ denotes the kinetic energy of a given pion in the $\eta$ rest frame, $Q = m_\eta-m_{\pip}-m_{\pim}-m_{\pio}$ is the kinetic energy in the reaction, and $m_{\eta/\pi}$ are the nominal masses from the PDG~\cite{PDG2014}. The decay amplitude of $\eta\to\pip\pim\pio$ can be parameterized as
\begin{equation}
  \begin{aligned}
        &|\mathcal{A}(X, Y)|^2\\
        &=N(1+aY+bY^2+cX+dX^2+eXY+fY^3+gX^2Y...),
  \end{aligned}
\end{equation}
where $N$ is a normalization factor and the coefficients $a, b, c, ...$ are called Dalitz parameters, when a non-zero value for $c$ or $e$ may imply the violation of charge conjugation.

Since no evidence of the charge-conjugation violation is seen in the previous measurement, we did not take this effect into account in the generator.   The parameters for $\eta\to\pip\pim\pio$ taken from the BESIII measurement~\cite{eta_3pi_bes} are:
\begin{equation}
  \begin{aligned}
        &a =-1.128,\\
        &b = 0.153,\\
        &d = 0.085,\\
        &f = 0.173.\\
  \end{aligned}
\end{equation}
%where the parameters $c$, $e$ and $g$ are set to be zero in the DIY generator.

For convenience, we also provide an option by including the item $X^2Y$ in the generator, and the parameters are from the KLOE-2 measurement~\cite{eta23pi_g}:
%The KLOE-2 Collaboration has also measured the Dalitz plot parameters with the contribution from $X^2Y$~\cite{eta23pi_g}
\begin{equation}
  \begin{aligned}
        &a =-1.095,\\
        &b = 0.145,\\
        &d = 0.081,\\
        &f = 0.141,\\
        &g =-0.044.\\
  \end{aligned}
\end{equation}

For $\eta/\eta'\to\pio\pio\pio$, the density distribution of the Dalitz plot has threefold symmetry, due to the three identical particles in the final states. Hence, the density distribution can be parameterized using the polar variable ~\cite{eta23pi0_amp},
\begin{equation}
  \begin{aligned}
        Z = X^2+Y^2 = \dfrac{2}{3}\sum_{i=1}^3(\dfrac{3T_i}{Q}-1)^2,
  \end{aligned}
\end{equation}
and the parametrization of the decay amplitude is given by~\cite{eta23pi_beta}
\begin{equation}
  \begin{aligned}
        |\mathcal{A}(Z)|^2 = N(1+2\alpha Z+2\beta Z^{3/2}\sin(3\phi)...),
  \end{aligned}
\end{equation}
where $T_i$ denotes the kinetic energies of each $\pio$ in the $\eta/\eta'$ rest frame, $Q=m_{\eta/\eta'}-3m_{\pio}$, $\phi = \arctan(Y/X)$,
and $N$ is the normalized factor.
$\alpha$ and $\beta$ are the Dalitz plot parameters. A nonzero $\alpha$ indicates final-state interactions.
$\beta$ has not been  measured yet, so it is set to zero in the generator, while the value of
$\alpha$ is taken from the BESIII measurement~\cite{eta_3pi_bes},
\begin{equation}
\alpha=
\left\{
\begin{aligned}
-0.055, ~\text{for} ~\eta\rightarrow\pio\pio\pio; \\
-0.640, ~\text{for} ~\eta'\rightarrow\pio\pio\pio. \\
\end{aligned}
\right.
\end{equation}

At the truth level,  the distributions of the Dalitz plot variables from the MC simulation  are shown in Fig.~\ref{eta23pi_draw}.

\end{multicols}
%\ruleup
\begin{center}
\includegraphics[width=14cm]{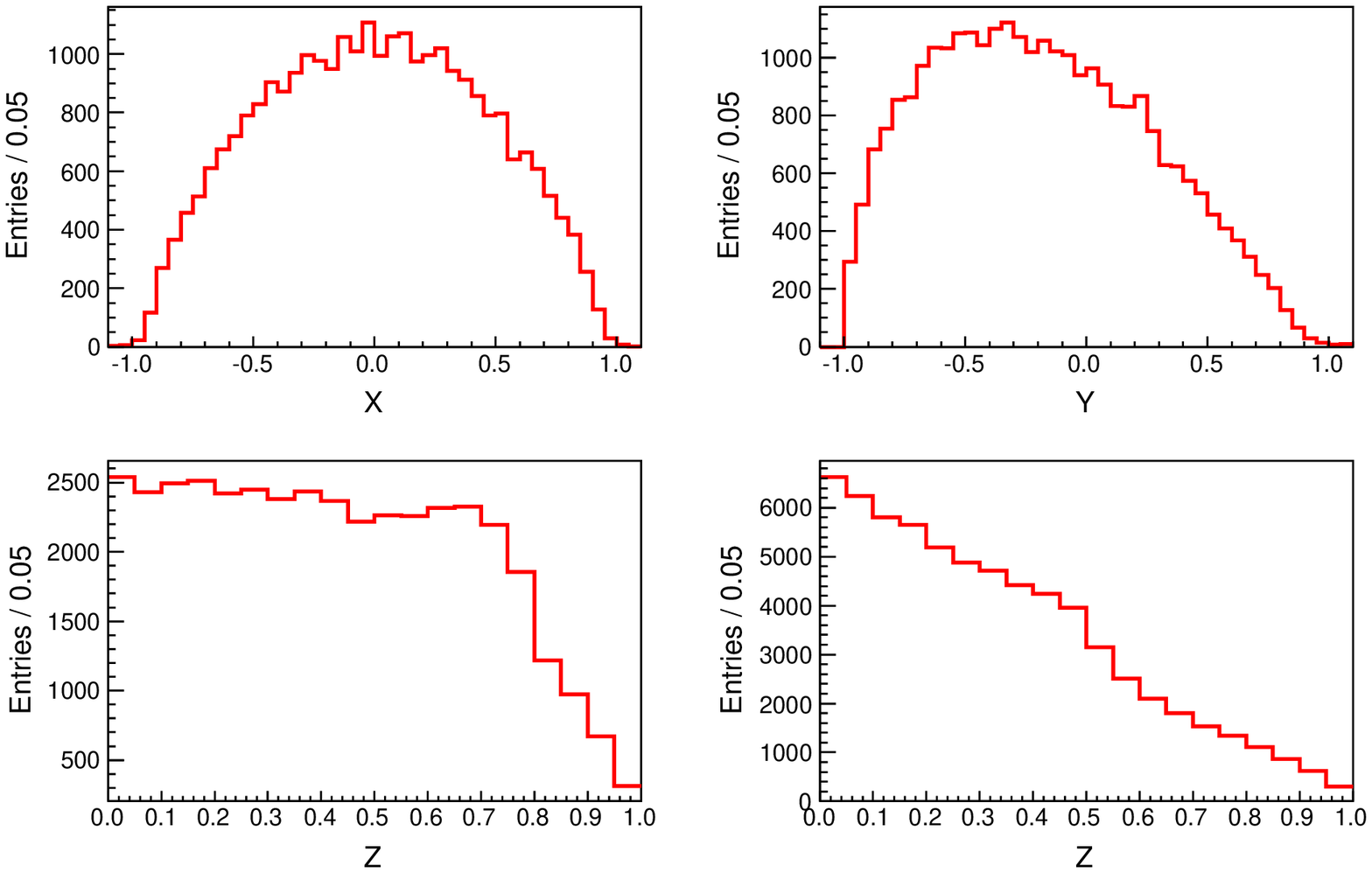}\put(-230,230){\bf (a)}\put(-30,230){\bf (b)}\put(-230,100){\bf (c)}\put(-30,100){\bf (d)}
\figcaption{\label{eta23pi_draw} The distributions of the Dalitz variables  (a) X  and (b)  Y from $\eta\to\pip\pim\pio$. (c) and (d) are the distributions of Z from $\eta\to\pio\pio\pio$ and $\eta'\to\pio\pio\pio$, respectively.}
\end{center}
%\ruledown

\begin{multicols}{2}

\subsection{$\eta'\to\eta\pi\pi$}
The matrix elements of $\eta'\to\eta\pi\pi$ have been measured by many experiments~\cite{etap2etapipi_ex}. The Dalitz plot for the charged channel $\eta'\to\eta\pip\pim$ can be described by two variables,
\begin{equation}
  \begin{aligned}
        &X = \dfrac{\sqrt{3}}{Q}(T_{\pip}-T_{\pim}),\\
        &Y = \dfrac{m_\eta+2m_\pi}{m_\pi}\dfrac{T_\eta}{Q}-1.
  \end{aligned}
\end{equation}
For the neutral channel $\eta'\to\eta\pio\pio$, due to the symmetry of the two $\pio$, the variable $X$ is replaced by
\begin{equation}
  \begin{aligned}
        X = \dfrac{\sqrt{3}}{Q}|T_{\pio_1}-T_{\pio_2}|,
  \end{aligned}
\end{equation}
where $T_\pi$ and $T_\eta$ are the kinetic energies of the mesons in the $\eta'$ rest frame and $Q=m_{\eta'}-m_\eta-2m_\pi$ is the kinetic energy in the decay, with $m_{\eta/\pi}$ the nominal masses in the PDG~\cite{PDG2014}.

For the general representation,  the squared amplitude is parameterized as
\begin{equation}
  \begin{aligned}
        |M(X, Y)|^2 = N(1+aY+bY^2+cX+dX^2+...),
  \end{aligned}
\end{equation}
where N is a normalization factor, and $a$, $b$, $c$ and $d$ are real parameters.
A non-zero $c$ parameter indicates violation of C parity for $\eta'\to\eta\pip\pim$ and violation of Bose symmetry for $\eta'\to\eta\pio\pio$.

An alternative parameterization is the so-called ``linear representation", which is written as follows:
%to parameterize the squared amplitude is call the ``linear parameterized'' one
\begin{equation}
  \begin{aligned}
        |M(X, Y)|^2 = N(|1+\alpha Y|^2+cX+dX^2+...),
  \end{aligned}
\end{equation}
where the complex parameter $\alpha$ can be compared with the general parameterization with $a=2Re(\alpha)$ and $b=Re^2(\alpha)+Im^2(\alpha)$. The real component of the complex constant $\alpha$ is a linear function of the kinematic energy of the $\eta$. The two parameterizations are equivalent in the case of  $b>a^2/4$.

In the case of the general representation, the parameters in the generator for  $\eta'\to\eta\pip\pim$ are taken from the BESIII measurement~\cite{etap2etapipi_ex}, which are:
\begin{equation}
  \begin{aligned}
        &a =-0.047,\\
        &b =-0.069,\\
        &c = 0.019,\\
        &d =-0.073.\\
  \end{aligned}
\end{equation}
These values are also used in the simulation for the neutral channel $\eta'\to\eta\pio\pio$.

With the above generator, the distributions from the MC simulation at the truth level are shown in Fig.~\ref{etap2pipieta_draw}.
The shapes of the Y variables in Fig.~\ref{etap2pipieta_draw}(b) and Fig.~\ref{etap2pipieta_draw}(d) are similar as a result of isospin symmetry.
There is a deviation between Fig.~\ref{etap2pipieta_draw}(a) and (c) due to different kinematics for X (no dynamics in the matrix elements).
The physical cusp effect was  not taken into account in the  generator because  no evidence has been observed yet~\cite{etap2etapipi_cusp}.
\end{multicols}
%\ruleup
\begin{center}
\includegraphics[width=14cm]{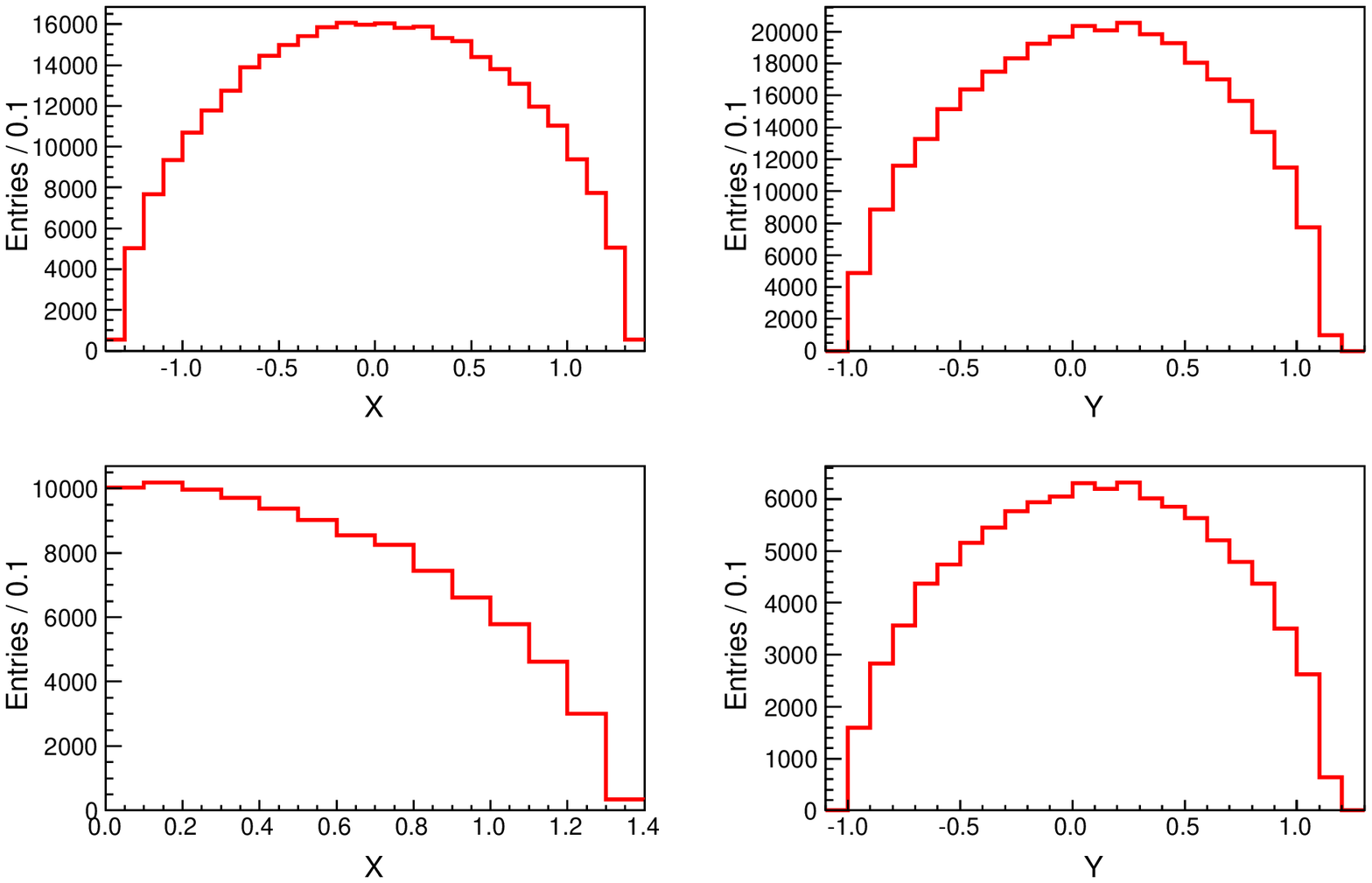}\put(-230,230){\bf (a)}\put(-30,230){\bf (b)}\put(-230,100){\bf (c)}\put(-30,100){\bf (d)}
\figcaption{\label{etap2pipieta_draw}
The distributions of the Dalitz plot variables (a) X and (b) Y from the MC simulation of $\eta'\to\eta\pip\pim$ events, and  (c) X  and (d) Y from the MC simulation of
$\eta'\to\eta\pio\pio$ events.}
\end{center}
%\ruledown

\begin{multicols}{2}

\subsection{$\eta'\to\pip\pim\pip\pim/\pip\pio\pim\pio$}
In ChPT, $\eta'\to4\pi$ is believed to be governed by the WZW term via chiral anomalies. Recently BESIII reported the first observation of $\eta'\to\pip\pim\pip\pim$ and $\eta'\to\pip\pim\pio\pio$ decays coming from $\jpsi\to\gamma\eta'$ ~\cite{etap_4pi_bes}. The measured branching fractions, $\BR(\eta'\to\pip\pim\pip\pim)=[8.53\pm0.69\pm0.64]\times10^{-5}$ and $\BR(\eta'\to\pip\pim\pio\pio)=[1.82\pm0.35\pm0.18]\times10^{-4}$, are consistent with the theoretical predictions based on a combination of ChPT and the VMD model~\cite{etap24pi_th2}.
% and the theoretical results are obtained to be $\BR(\eta'\to\pip\pim\pip\pim)=(1.0\pm0.3)\times10^{-4}$, $\BR(\eta'\to\pip\pim\pio\pio)=(2.4\pm0.7)\times10^{-4}$. BESIII reported the first observation of $
%\eta'\to\pip\pim\pip\pim$ and $\eta'\to\pip\pim\pio\pio$ decays coming from $\jpsi\to\gamma\eta'$ in 2014~\cite{etap_4pi_bes}. The measured branching fractions are $
%\BR(\eta'\to\pip\pim\pip\pim)=[8.53\pm0.69\pm0.64]\times10^{-5}$ and $\BR(\eta'\to\pip\pim\pio\pio)=[1.82\pm0.35\pm0.18]\times10^{-4}$, which are consistent with theoretical predictions. The decay amplitudes in the DIY %event generators are based on Ref.~\cite{etap24pi_th2}.

By following the notations in Ref.~~\cite{etap24pi_th2}, the four-momenta are defined as

\begin{equation}
  \begin{aligned}
        &\eta'\to \pip(p_1)\pim(p_2)\pip(p_3)\pim(p_4),\\
        &\eta'\to \pip(p_1)\pio(p_2)\pim(p_3)\pio(p_4).
  \end{aligned}
\end{equation}
The amplitudes can be described in terms of the invariant variables $s_{ij}=(p_i+p_j)^2, i,j=1,...,4$, which are subjected to the constraint (in the isospin limit of equal pion masses)
\begin{equation}
  \begin{aligned}
        s_{12}+s_{13}+s_{14}+s_{23}+s_{24}+s_{34}=m^{2}_{\eta'}+8m^{2}_{\pi}.
  \end{aligned}
\end{equation}
%The five-meson vertices of the WZW term can be deduced from the Lagrangian
%\begin{equation}
%  \begin{aligned}
%        {\cal L}^{WZW}_{P^5}=\dfrac{N_c\epsilon_{\mu\nu\alpha\beta}}{240\pi^2f^5_{\pi}}\langle\varphi\partial^{\mu}\varphi\partial^{\nu}
%        \varphi\partial^{\alpha}\varphi\partial^{\beta}\varphi \rangle+...,
%  \end{aligned}
%\end{equation}
%where $\langle\varphi\partial^{\mu}\varphi\partial^{\nu}\varphi\partial^{\alpha}\varphi\partial^{\beta}\varphi\rangle$ represents the trace in flavour space, $f_{\pi}=92.4$MeV is the pion decay constant, and $N_c=3$ is the number of colors. Meanwhile,
%\begin{equation}
%  \begin{aligned}
%        \dfrac{\varphi}{\sqrt2}=\begin{pmatrix}
%                                    \dfrac{\eta_0}{\sqrt3}+\dfrac{\eta_8}{\sqrt6}+\dfrac{\pi_0}{\sqrt2}&\pip&K^{+}\\
%                                    \pim&\dfrac{\eta_0}{\sqrt3}+\dfrac{\eta_8}{\sqrt6}-\dfrac{\pi_0}{\sqrt2}&K^{0}\\
%                                    K^{-}&\bar{K}^{0}&\dfrac{\eta_0}{\sqrt3}-\dfrac{2\eta_8}{\sqrt6}
%                                 \end{pmatrix}.
%  \end{aligned}
%\end{equation}
The $\eta-\eta'$ mixing is described as
\begin{equation}
  \begin{aligned}
        |\eta\rangle&=\cos\theta_{mix}|\eta_8\rangle-\sin\theta_{mix}|\eta_0\rangle,\\
        |\eta'\rangle&=\sin\theta_{mix}|\eta_8\rangle+\cos\theta_{mix}|\eta_0\rangle.
  \end{aligned}
\end{equation}
The mixing angle is $\theta_{mix}=-20^{\circ}$~\cite{parameters_s1}.

The decay amplitudes are then given by
\begin{equation}
  \begin{aligned}
        &\mathcal{A}_V(\eta_8\to\pip\pim\pip\pim) = \dfrac{1}{\sqrt2}\mathcal{A}_V(\eta_0\to\pip\pim\pip\pim)\\
        = &-\mathcal{A}_V(  \eta_8\to\pip\pio\pim\pio )= -\dfrac{1}{\sqrt2}\mathcal{A}_V(  \eta_0\to\pip\pio\pim\pio  )\\
        = &\dfrac{N_c\epsilon_{\mu\nu\alpha\beta}}{16\sqrt3\pi^2f^5_{\pi}}p^{\mu}_1p^{\nu}_2p^{\alpha}_3p^{\beta}_4
        \left\{\left[\dfrac{m^2_{\rho}}{D_{\rho}(s_{12})}\right.\right.+\dfrac{m^2_{\rho}}{D_{\rho}(s_{34})}-\dfrac{m^2_{\rho}}{D_{\rho}(s_{14})}\\
        &-\left.\dfrac{m^2_{\rho}}{D_{\rho}(s_{23})}\right](c_1-c_2-c_3)+2c_3\left[\dfrac{m^4_{\rho}}{D_{\rho}(s_{12})D_{\rho}(s_{34})}\right.\\
        &-\left.\left.\dfrac{m^4_{\rho}}{D_{\rho}(s_{14})D_{\rho}(s_{23})}\right]\right\},
  \end{aligned}
\end{equation}
where $f_{\pi}=92.4$ MeV is the pion decay constant~\cite{parameters_s1}, and $N_c=3$ is the number of colors.
\begin{equation}
  \begin{aligned}
        D_{\rho}(s)&=m^2_{\rho}-s-im_{\rho}\Gamma(s)
  \end{aligned}
\end{equation}
is the inverse $\rho$ propagator.
$\Gamma(s)$ is the same as in Eq.~(\ref{eq5_gamLL}).
The coupling constants are assigned to be $c_3=c_1-c_2=1$.

The mass spectra of $\pi\pi$ at the truth level  are shown in Fig.~\ref{etap24pi_draw} and Fig.~\ref{etap24pi0_draw}. For the generator of $\eta^\prime\to\pi^+\pi^-\pi^+\pi^-$,
it has been shown in Ref.~\cite{etap_4pi_bes} that the simulated $\pi^+\pi^-$ invariant mass distribution is more consistent with data than that from the uniform phase space events.
For the generator of $\eta^\prime\to\pi^+\pi^-\pi^0\pi^0$, it has not been validated in Ref.~\cite{etap_4pi_bes} due to the limited statistics of the data.
%{\color{red}
%The full simulations of $\eta'\to4\pi$ with these generators considering the BESIII detector performance is consistent with the background-subtracted data~\cite{etap_4pi_bes}.
%}
%\ruleup
\begin{center}
  \includegraphics[width=7cm]{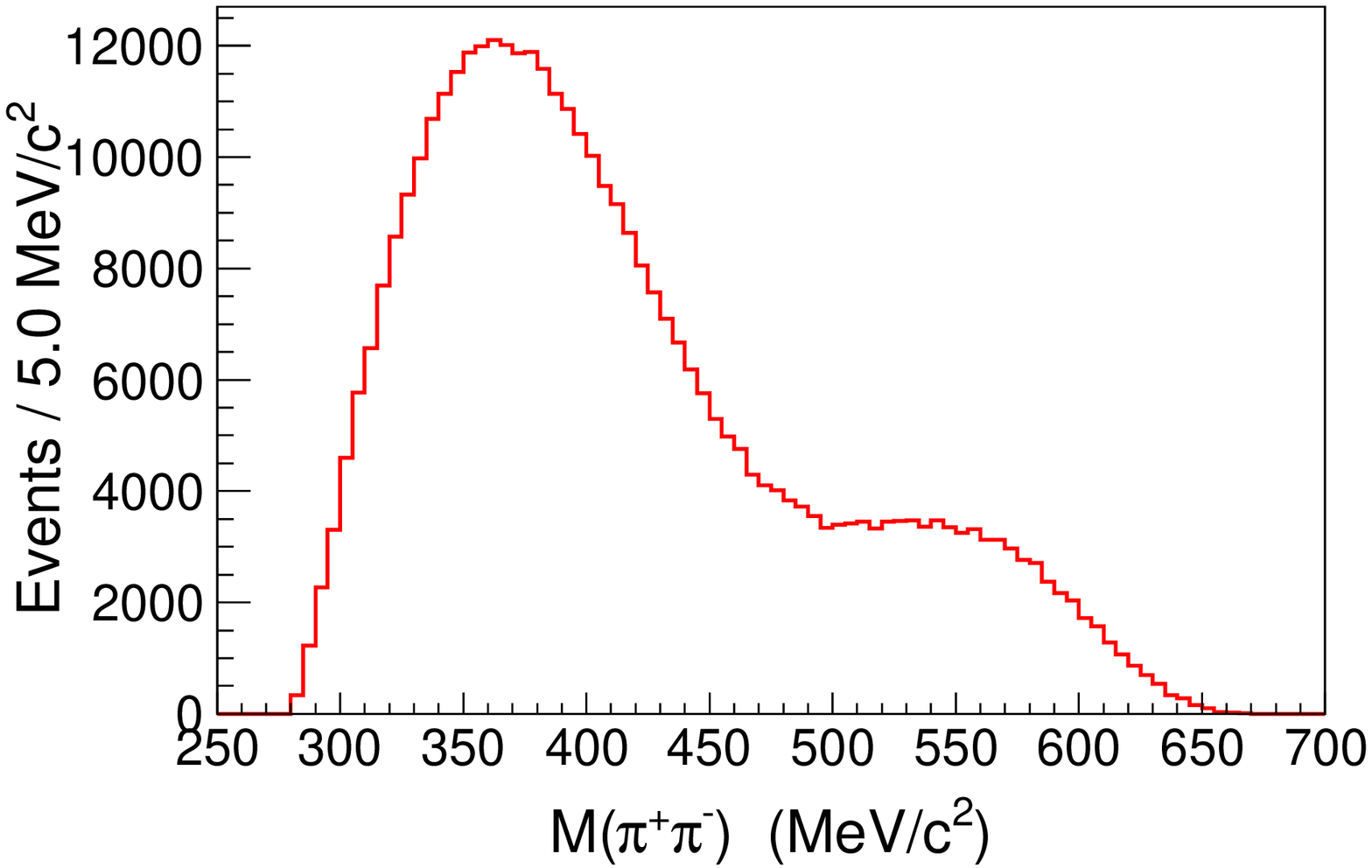}
  \figcaption{\label{etap24pi_draw}The invariant mass spectrum of $\pip\pim$ from $\eta'\to\pip\pim\pip\pim$(4 entries per event).}
\end{center}

\end{multicols}
%\ruleup
\begin{center}
\includegraphics[width=14cm]{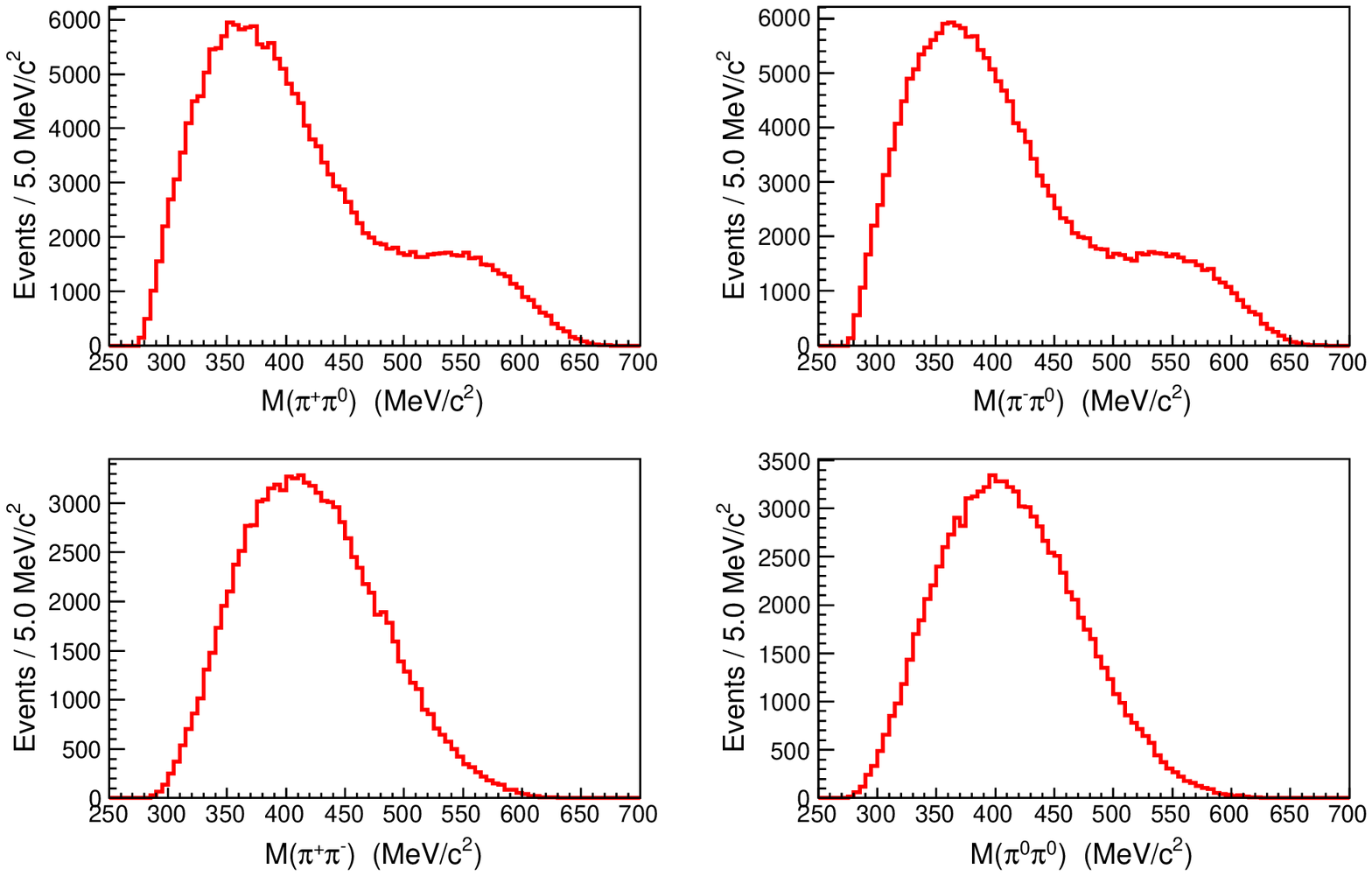}\put(-230,230){\bf (a)}\put(-30,230){\bf (b)}\put(-230,100){\bf (c)}\put(-30,100){\bf (d)}
\figcaption{\label{etap24pi0_draw}The invariant mass distributions  of (a) $M(\pi^+\pi^-)$, (b) $M(\pi^-\pi^0)$, (c) $M(\pi^+\pi^-)$ and (d) $M(\pi^0\pi^0)$ from the simulated $\eta'\to\pip\pim\pio\pio$ events, where there are 2 entries per event in (a) and (b).}
\end{center}
%\ruledown

\begin{multicols}{2}

\section{Summary}
Based on the amplitudes of $\eta/\eta'$ decays calculated by ChPT and the VMD model,  we have developed a series of event generators for $\eta/\eta^\prime$ decays, including $\eta/\eta^{\prime}\to\gamma l^{+}l^{-}$, $\eta/\eta^{\prime}\to\gamma \pi^{+}\pi^{-}$, $\eta^{\prime}\to\omega e^{+}e^{-}$, $\eta\to\pi^{+}\pi^{-}\pi^{0}$, $\eta/\eta'\to\pio\pio\pio$, $\eta^{\prime}\to\eta\pi\pi$ and $\eta'\to\pip\pim\pip\pim/\pip\pim\pio\pio$.   Most of them have been validated in the study of $\eta/\eta^\prime$ decays at BESIII and the parameters tuned accordingly to provide a good description of data. Indeed, these event generators play an important role in the observations of new decay modes of the $\eta^\prime$ meson~\cite{etap_4pi_bes, etap_omegaee_bes} reported by the BESIII Collaboration.

At present, the world's largest sample of $J/\psi$ events, $1.31\times10^9$ events collected with the BESIII detector, provides a unique opportunity to investigate $\eta/\eta'$ decay dynamics. Besides the achievements obtained from the $J/\psi$ radiative or hadronic decays into $\eta/\eta^\prime$~\cite{etap_pipiee_bes, etap_gammaee_bes, eta_3pi_bes, etap_4pi_bes, etap_omegaee_bes, etap_Kpi_bes}, many analyses on $\eta/\eta^\prime$ physics are in progress, which offer an opportunity to further evaluate the usability of these generators by examining whether they can provide a good description of data.  In addition to the BESIII experiment, these event generators could also be a useful tool to investigate $\eta/\eta^\prime$ decays in other experiments, e.g., GlueX, CLAS12, and KLOE-2.

\vspace{5mm}
\acknowledgments{Nian Qin thanks Dr. Xiao-Lin Kang, Dr. Xin-Ying Song, Dr. Li-Qing Qin and Ms. Hui-Juan Li for their support in developing the generators and helpful discussions.}

\end{multicols}

\vspace{-1mm}
\centerline{\rule{80mm}{0.1pt}}
\vspace{2mm}

\begin{multicols}{2}

\end{multicols}

\clearpage

\end{CJK*}
\end{document}